# Posterior corrections for systematic distortions in atomic-resolution images from hexagonal crystals*

T. Bortel and P. Moeck

*Abstract*—Digital images from crystals, as projected from the third spatial dimension and recorded in atomic resolution with any kind of real-world microscope, feature necessarily broken symmetries of the translation-periodicity-restricted Euclidean plane. The symmetry breakings are due to both the imaging process and the real structure of the imaged crystal, with the former cause typically dominating. A posterior algorithmic reduction of the symmetry breaking in such images constitutes, thus, often a correction for many of the distortions that were introduced by the imaging processes. Numerically quantified restorations of such symmetries can, therefore, be used to demonstrate the efficacy of a newly implemented posterior correction method for atomic-resolution images from hexagonal crystals. Recently developed information theory based methods are here shown to be suitable for this purpose. Thirteen experimental atomic-resolution images from graphite and monolayer molybdenite ($MoS_2$), as respectively obtained by scanning tunneling microscopy, atomic force microscopy in the torsional resonance mode, and aberration-corrected parallel-illumination transmission electron microscopy served as test cases in our larger (in its totality so far unpublished) study, from which we quote here. The source code of the software that was used for our distortion corrections and the whole report on that study are freely available on GitHub.

## I. Introduction

When capturing atomic resolution imagery, such as that recorded by various types of scanning probe microscopy (SPM) and aberration-corrected transmission electron microscopy (AC-TEM) at very low acceleration voltages, systematic imaging errors are not unusual. In standard scanning tunneling microscopy (STM) for example, a conducting sample or very fine metallic probing tip is moved relative to the other by piezoelectric actuators. These actuators have frequently the shape of a tube and allow for both sample movements up to hundreds of nanometers for selecting scanning areas of interest and much higher precision movements for scanning in the nanometer range for the actual imaging at the chosen location on the sample. This necessitates relatively large bending of the tube to scan peripheral sample areas, resulting in nonlinear piezoelectric responses, hysteresis, creep, and instrumental drift, which is often exacerbated by small temperature changes within parts of the SPM apparatus. The scanned area of the sample thus often has the shape of a parallelogram, whereas a square or rectangular image area is displayed to the unsuspecting operator of the microscope and recorded in the image, possibly leading to some *"untenable conclusions"* [1] about a sample being drawn from compromised image data.

We present here a methodology and our computer program for correcting these (and other) sorts of imaging errors in atomic resolution images based on ideas in [1]. Objective crystallographic symmetry classifications [2] of such images into plane symmetry groups and projected Laue classes are here used to demonstrate the efficacy of our newly implemented method in combination with measures of broken lattice symmetries. Whereas there are several alternatives to the approach taken in [1] and some alternative software implementations to be found in the literature, e.g. [3], their efficacy in restoring systematically broken crystal symmetries in corrected images has to the best of our knowledge not been demonstrated before. The reason for this has quite obviously been the lack of objective plane symmetry classification methods [4] until the quite recent appearance of [2].

The organization of the rest of this paper is as follows. Firstly, the objective symmetry classification methods used to evaluate the efficacy of our correction method are discussed briefly at the beginning of the background section. Highly oriented pyrolytic graphite (HOPG) is discussed as a calibration and support sample for SPM at the end of the background section. Secondly, the mathematical underpinnings of our distortion and correction models are briefly introduced, and our MATLAB code implementing them is described in the subsequent two sections. That code can be freely downloaded from GitHub [5]. Thirdly, three atomic-resolution STM images of HOPG are used to demonstrate the effectiveness of our correction method. Finally, the results of testing our posterior image correction method on eleven other atomic-resolution images (including a single AFM image from HOPG, two experimental AC-TEM images from monolayers of molybdenite, and a single synthetic image that resembles HOPG) are briefly summarized.

## II. Background

### A. Objective Crystallographic Symmetry Classifications

Historically, determining which of the 17 plane symmetry groups a particular crystal pattern in two dimensions (2D) belongs to has been an essentially subjective process [4]. We use the words "crystal pattern" here in the crystallographic sense [6], which frees us from giving experimental imaging details below as they are utterly unimportant for both our objective symmetry classifications [2] and our image-distortion corrections [7], that effectively remove projected centro-symmetry that has been imposed by the imaging process itself. Note that all symmetries in experimental atomic-resolution images from crystals are always broken to a larger or smaller extent. Thus, they constitute noisy and, more often than not, systematically distorted 2D crystal patterns.

*Research supported by Portland State University and a direct access grant from the Oak Ridge National Laboratory.

T. Bortel is a graduate student at the Department of Physics of Portland State University, Portland, OR 97207-751, USA, phone: 415-686-3475, email: tylerbortel@gmail.com.

P. Moeck is an emeritus professor in the same department, phone: 503-725-4227, e-mail: pmoeck@pdx.edu.

Arbitrarily set thresholds have been used in the past to aid in subjectively determining if a certain symmetry or combination of symmetries are statistically present in a 2D crystal pattern or not. There was, until recently, also no way to deal objectively with pseudo-symmetries of the Fedorov type [2].

Novel methods have, however, been recently demonstrated to make objective classifications of crystallographic symmetries using a form of geometric information theory as their basis [2]. The decision thresholds in those methods are not arbitrarily set but derive from both the hierarchy of symmetries of 2D crystal patterns and quantified amounts of symmetry breakings under the assumption that (unavoidable) random imaging errors are more or less Gaussian distributed. The quantitative backbone of this classification method is sums of squared residuals that are calculated by comparing a raw-data crystal pattern with all pertinent symmetry-enforced versions of it, conveniently done in Fourier space [8].

For a hexagonal 2D crystal pattern, the amplitude map of the discrete Fourier transform (DFT) yields sets of six peaks, situated more or less equidistantly around the origin of the map in three pairs. There are three possible choices of primitive reciprocal axes, one for each pair of non-colinear peaks, and each is simultaneously compatible with three choices of centered reciprocal axes. Given a selection of reciprocal axes and data extracted from a Fourier transform, using a dedicated electron crystallography program such as CRISP [9], sums of squared residuals can be calculated for each of the 17 plane symmetry groups in all applicable 21 settings [2, 8].

With such sums of residuals calculated, one makes a determination of the plane symmetry group that the crystal pattern most likely possesses in a statistic/probabilistic sense. Because of the subgroup-supergroup nature of the plane symmetry groups, classifications are made by first identifying the presence of simpler and less broken symmetries, and then "climbing up in the symmetry hierarchy", if possible, to larger symmetry groups that are (translationengleiche [10]) *minimal* supergroups thereof.

The classification procedure begins with the selection of an anchor group, which is taken as the plane symmetry group with the lowest sum of squared complex Fourier coefficient residuals that has either two or three non-translational symmetry operations. This means that the anchoring group will be either *p2*, *p3*, or one of the *pg*, *cm*, or *pm* plane symmetry groups. To determine if one is objectively justified to "climb" to a higher-level symmetry which is a minimal supergroup of the anchor group, the ratio of the sums of squared residuals is compared to its maximal allowed value to conclude if the higher symmetry is statistically present

$$\frac{\hat{J}_m}{\hat{J}_l} < 1 + \frac{2\left(k_m - \left(\frac{N_m}{N_l}\right)k_l\right)}{k_m(k_l - 1)} \quad (1),$$

where $k_m$ and $k_l$ are the number of non-translational symmetry operations present in the more symmetric and less symmetric plane symmetry groups, $N_m$ and $N_l$ are the number of data points used in the measurement of their sums of respective squared residuals, and $\hat{J}_m$ as well as $\hat{J}_l$ are the calculated sums of squared residuals (where the subscript *m* stands for *m*ore, subscript *l* stands for *l*ess, and the tiny hat signifies that they are both statistical estimates which are directly connected to the experimental image itself [2, 8]).

An experimentally obtained ratio of the sums of squared residuals of one in inequality (1) indicates 100% confidence in a particular ascent to the respective minimal supergroup. The right-hand side of inequality (1) varies from 1.2 to seven-thirds for different supergroup and subgroup combinations (in cases of $N_m = N_l$). Whenever the ratio on the left-hand side of this inequality is larger than these values, the confidence of the classification into a respective minimal supergroup drops to zero, i.e. there is no longer a statistically sound justification for considering that higher symmetric group as being the better statistical representation of the translation periodic information in the crystal pattern.

In order to allow for a "symmetry climb", the *minimal* supergroup must be connected through a series of allowed upward climbs starting at the anchor group, and it must be possible to climb from all *maximal* subgroups. If these criteria cannot be met, then the climb cannot be made, and the lower symmetric group, which might be the anchor group itself, is statistically best supported by the image data. Confidence levels as defined in [2] are considered to be a form of probability so that the product of the confidence levels of all anchored and allowed climbing steps upwards in the symmetry hierarchy tree are multiplied in order to give the total confidence of a classification.

In addition to the plane symmetry group, a related series of operations yields the projected Laue class of a 2D crystal pattern [2]. Much like for plane symmetry group classification, the projected Laue class tree is climbed up from the anchor class when an analogue of inequality (1) is fulfilled for each *maximal* subgroup toward a particular *minimal* supergroup for the translation periodicity restricted 2D point symmetries that are centro-symmetric, i.e. that contain a two-fold rotation point.

These two classifications (plane symmetry group and projected Laue class) are always compatible with each other for synthetic image data [7] that are only subject to random imaging errors which are more or less Gaussian distributed. For experimental image data, one often runs into a crystallographic inconsistency that reflects the presence of systematic imaging distortions.

In addition to the objective symmetry classifications and associated confidence levels, two composite Bravais lattice distortion metrics are used here. They assess the "lattice symmetry faithfulness" of the images both before and after correction. "Angle Deviation" refers to the standard deviation of the three angles measured between the various choices of primitive axes in Fourier space. Even if a correction is not sufficient to allow for an improvement in the symmetry group and/or projected Laue class classification from a known sample, or if the original image already features the correct symmetry, this metric captures some indication of the improvement due to the correction. The same is true for the second composite metric, "Mean % Length Difference" which calculates the average difference between the length of the reciprocal axes as a percentage of that length.

## B. Typical image distortions in scanning probe microscopy and HOPG as a support/calibration sample

There are a multitude of factors that can lead to distortions in images from any kind of microscopy. Some of these distortions are localized while others are global. Of particular interest to the various SPM technologies are distortions due to instrumental drift, which is quite ubiquitous. To the first order, these distortions can be modeled by a 2D vector [1], which amounts to a linear transformation of the scanning area from its intended square or rectangle to a parallelogram as shown in Fig. 1.

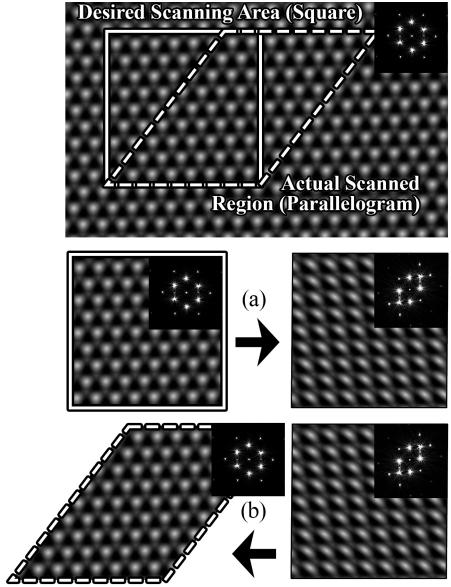

Figure 1. Visual representation of a global linear shear distortion (a) and its correction (b) with DFT amplitude map insets. *While the microscope "produces" a square image, it actually scans a parallelogram resulting in the systematic image distorting effects show on the right hand side of subfigures (a) and (b). Reversing this distortion results in a parallelogram image depicting the original area scanned as in (b).*

Our goal is to recreate, as accurately as possible, the actual parallelogram scanned by the imaging process, taking only the apparently resulting square or rectangular image as input. The *p3m1* crystal pattern shown here was created in Photoshop and resembles a STM image of HOPG with the second atom of the unit cell, i.e. the one in the first sub-surface layer, barely visible. As most SPM instruments do not record positional drifts of the sample or probing tip independently and with very high accuracy, they simply record and display images in the shape of a square or rectangle as illustrated in Fig. 1 although the imaged area was actually a parallelogram.

The geometric distortion correction method presented here is reliant on a trigonal/hexagonal symmetry structure in a 2D projection along the three-fold symmetry axis, which HOPG and molybdenite possess. HOPG and $MoS_2$ are both known to possess plane symmetry group *p3m1*, with *c1m1* and *p3* as *maximal* subgroups, which posses accompanying projected Laue classes of *6mm*, *2mm*, and *6*, respectively.

The ubiquity of these kinds of calibration/support samples lends itself to the following potential application of our correction method. In material sciences, pharmaceutical studies, and other disciplines that utilize atomic-resolution imagery from SPM, molecules and atomic clusters are often carefully imaged for improved understanding. With the rather pervasive presence of the above-mentioned systematic distortion in atomic-resolution imaging in SPM, it can be difficult to trust these images for use in quantitative measures such as atomic spacings, structural angles, and image contrasts. By positioning a specimen on a calibration sample such as a HOPG surface, symmetry measurements taken of the crystalline background can be used to apply our geometric correction method to the entire image. This corrected image would at the very least provide a more accurate measure of lengths, structural angles, and contrasts involved in the observation of the molecule or atomic cluster. Furthermore, the crystalline background could be removed from the image of the molecule or atomic cluster by Fourier filtering.

## III. THE IMPLEMENTED CORRECTION METHOD

As mentioned above, SPM scans often image an area incongruous with their intention, Fig. 1. We model this discrepancy in the form of a singular "distortion vector" that linearly transforms the intended square/rectangle scanning area to the parallelogram that was actually scanned. Our endeavor is to quantify this transformation in Fourier space, derive its inverse, and apply that inverse transformation to our original (direct space) image. The following is a summary of the mathematical underpinnings of the correction method as originally outlined by Henriksen and Stipp [1], but for which those authors did not provide a software implementation.

The linear transformation (L) has the form of a 2×2 matrix, with each column consisting of the transformed Cartesian unit vectors (1, 0) and (0, 1) respectively. As the global distortion is assumed to operate mainly along the y-axis (since drift parallel to the horizontal direction of the fast scan is much less likely due to the high speed of the scan in that direction), the result is the identity matrix plus the drift vector d′, which is (d′$_x$, d′$_y$),

$$L = \begin{pmatrix} 1 & 0 \\ 0 & 1 \end{pmatrix} + \begin{pmatrix} 0 & d'_x \\ 0 & d'_y \end{pmatrix} = \begin{pmatrix} 1 & d'_x \\ 0 & 1 + d'_y \end{pmatrix} \quad (2).$$

In Fourier space, this linear transformation distorts the ideally presumed circle defined by six periodic-structure bearing peaks in the DFT amplitude map that are symmetry equivalent around its center into an ellipse. Choosing three non-colinear peaks a = ($a_x$, $a_y$), b = ($b_x$, $b_y$), and c = ($c_x$, $c_y$) in the amplitude map of the DFT, we can then describe this ellipse with three equations of the form:

$$k^2 \left[ (a_x + q_x a_y)^2 + (q_y a_y)^2 \right] = 1 \quad (3),$$

where *k* is a scale factor used to normalize the radius of the original circle, $q_x$ = d′$_x$, and $q_y$ = 1 + d′$_y$. Some algebra yields

$$q_x = \frac{1}{2} \frac{a_y^2 b_x^2 - a_x^2 b_y^2 - a_y^2 c_x^2 + a_x^2 c_y^2 - b_x^2 c_y^2}{a_y^2(c_x c_y - b_x b_y) + b_y^2(a_x a_y c_x c_y) + c_y^2(b_x b_y - a_x a_y)} \quad (4),$$

$$q_y = \sqrt{\frac{a_x^2 + 2 a_x a_y q_x + a_y^2 q_x^2 - b_x^2 - 2 b_x b_y q_x - b_y^2 q_x^2}{b_y^2 - a_y^2}} \quad (5).$$

The Cartesian positions of three points along the ellipse are thus sufficient information from which to derive the modeled distortion vector as illustrated in Fig. 2.

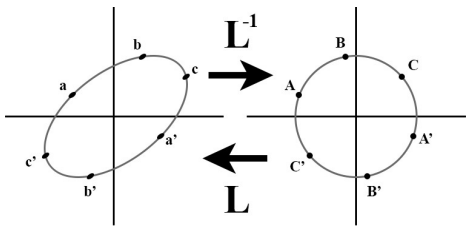

Figure 2. Sketch depicting a shear correction in the amplitude map of a Fourier transform (DFT). *Pictured before (left) and after (right) correction. The three chosen peaks a, b, and c, are used to determine the linear transformation (to peaks A, B, and C) and its inverse, which is to be applied in direct space. The peaks a', b', and c' could be used alternatively.*

Objective symmetry classifications [2, 8] compliment and enhance such a geometric correction for systematic image distortions. By having available quantitative measures of the efficacy of an implemented correction, the correction method may be applied iteratively. If, after a correction, an image still fails to register with the known symmetry of the imaged crystal and/or other image metrics are still deemed unsatisfactory, the operator can simply make the peak selection of Fig. 2 in the amplitude map of the DFT of the corrected image to facilitate an additional round of correction, as sketched out in Fig 3. This has been shown, as in the second test case below, to be an effective strategy when working with images suffering from severe systematic distortions.

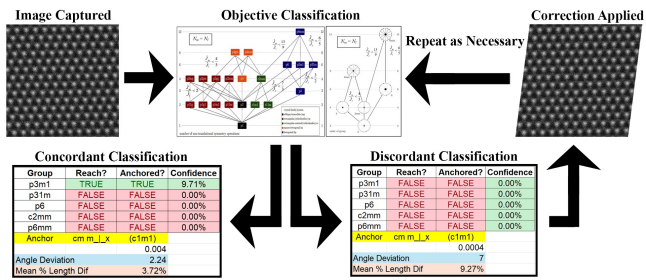

Figure 3. Graphic abstract detailing the iterative processes of Symmetry Classification and Distortion Correction. *The classification trees in the top row within the Objective Classification window were taken from [2] and represent the applicable subgroup/supergroup relationships. The tables in the bottom row are parts of plane symmetry classification results in the exel format, as detailed in [7].*

## IV. IMPLEMENTATION AND MATLAB CODE

The MATLAB software that implements this method runs in two phases [7]. The first phase takes as input the amplitude map of the discrete Fourier transform of the image to be corrected and extracts the necessary information to decipher the transformation necessary to correct that original image. The second phase calculates and applies the necessary transformation, returning a "corrected" image file.

### A. Program - Main1

The first of the two routines that needs to be run to correct an image is "Main1". This program first prompts the user to select the DFT amplitude map of the original image. (This map can be calculated with a program such as CRISP). The first "rough" peak finding is performed by a pre-made algorithm developed by Adi Natan [11]. Once the map has been uploaded, the user is prompted with a window containing the amplitude map of the DFT with all identified peaks highlighted with a "+" sign. Depending on the contrast in the uploaded map, there may only be a few peaks highlighted or there may be several others as well.

The sensitivity of the peak finding is set by default to be such that it only registers peaks at least 15% the intensity of the brightest pixel in the DFT amplitude map. This is controlled by the variable "minpeak" and can be freely altered should the desired peaks be insufficiently intense to register. With the peaks identified, it is then up to the user to determine which 3 peaks $(a_x, a_y)$, $(b_x, b_y)$, and $(c_x, c_y)$ will be used to calculate the distortion correction. Depending on the resolution in the original image, there may be more than one ring of peaks from which to choose. Each peak can be selected by clicking on it, which will reveal its coordinates.

### B. Program - Correct_Original_Image

The second routine in the sequence is "Correct_Original_Image". This program collects the original image to be corrected as well as the rough peak positions measured in "Main1" and outputs the corrected image as a ".jpg" file.

Upon running the program, the user is prompted to select the original image file to be corrected, before being provided a form to input the coordinates of the three pertinent peaks, see Fig. 2, found in "Main1". After all inputs have been submitted, the first step is to refine the rough peak locations. As the images used are often quite small, the singular pixel best approximating the peak is an insufficient degree of specificity: a sub-pixel peak location must be found. The algorithm for this is a 2D adaptation of the method outlined in [12], there entitled "CoM7" (short for Center of Mass: 7 Pixels). This routine assumes that our peak fits a Gaussian profile and offsets each peak from its rough location by a vector calculated from the values of a 7 x 7 pixel grid centered at the position of the rough peak.

With the coordinates of refined peaks relative to the origin calculated, the program is then able to compute the necessary linear transformation to closest map the ellipse formed by the three non-collinear Fourier-amplitude peaks to a circle with the same origin as described above. The program prompts the user to select the original (direct space) image to which the transformation is to be applied, the correction is made, and the corrected image is displayed (as well as saved automatically as a ".jpg" file).

## V. SELECTED TEST CASES

Each of the three test cases below will be introduced with their plane symmetries analyzed/classified in the original form of the atomic-resolution image. Then the post-correction version of the same 2D crystal pattern will be presented and its plane symmetries analyzed/classified. The STMs used for these three test cases feature piezo-electric tubes as actuators and are, thus, prone to instrumental drift effects as outlined at the beginning of the introduction.

## A. Test Case 1

This particular HOPG sample image, while not of the highest quality, is of sentimental value to the authors. It was recorded by them at the NanoTransport STM system of the Oak Ridge National Laboratory in 2022 and was a catalyst for this project. As shown in Fig. 4a, before the correction, this image features plane symmetry group *c1m1*, projected Laue class *2mm*, an Angle Deviation of 4.02°, and a mean reciprocal lattice vector Length Difference of 7.33%.

The corrected image, Fig. 4b, yields the predicted plane symmetry result of *p3m1* with the corresponding projected Laue class *6mm*, while reducing the Angle Deviation to an impressive 0.14°, and a mean Length Difference of only 0.29%. This serves as "anecdotal evidence" that even an image with apparently low visual quality may yield impressive correction results.

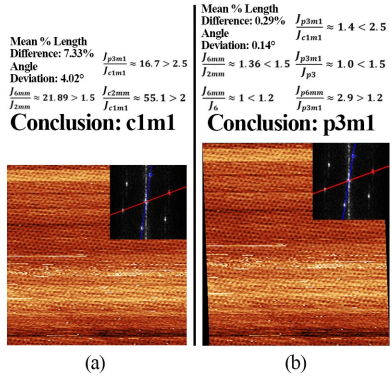

Figure 4. STM image of HOPG (Test Case 1). *Pictured before (a) and after (b) correction with DFT amplitude map insets, accompanied by respective ratios of J values and composite lattice distortion metrics. (Length scales have been dropped as atomic resolution is presented.)*

## B. Test Case 2

This STM image from another HOPG sample was recorded by somebody else at the same STM (at the Oak Ridge National Laboratory). While it seems to be of much higher visual "fidelity", see Fig. 5a, than the previous crystal pattern, it exhibits significantly stronger distortions than the image we took ourselves, Fig. 4a. The systematic distortions in Fig. 5a are so severe that the CRISP software used as part of the symmetry classifications was only able to detect one of the three sets of primitive axes and was unable to detect the possibility of centered lattice alternatives before correction. (As CRISP is a dedicated electron crystallography program, it was not designed to allow for the Fourier-space indexing of very severely distorted 2D crystal patterns [9].)

The singular choice of primitive reciprocal axes provided yields a substantial reciprocal vector Length Difference of 24.87%. With only one set of reciprocal axes, there can be no Angle Deviation, but that set has a measure of 31.7° for the angle between both reciprocal axes, which is significantly less than the hexagonal value of 60°. The plane symmetry displayed registers as *p2*, which is not a subgroup of plane symmetry *p3m1*, i.e. the projected symmetry of HOPG.

After correction, CRISP detected a centered lattice, but the plane symmetry only registers as *c1m1* (instead of *p3m1*), Fig. 5b. Note that the atoms appear visually "rather round" in this subfigure, and have typically six "nearest" neighbors (because any "three-fold" contrast due to atoms in the first subsurface layer at position ($^1/_3$,$^1/_3$) of the projected unit cell is too week to be readily seen), all as it should be. The mean % Length Difference falls significantly to 2.13%, and the reciprocal lattice angle measures 53.7° after correction. We are, however, not yet finished with that image. To improve the image quality and symmetry faithfulness even further, a circular-disk selection of the corrected image was made and run through the correction algorithm for a second time. The results of this are, as shown in Fig. 5c, a significant improvement over the singular correction. The objective symmetry classification reaches the full *p3m1*, with consistent projected Laue class *6mm*. While there is still only one detectable choice for primitive axes, the measure of the angle between them is even closer to 60 degrees, at 64.8°. The % Length Difference does rebound somewhat to 10.2%, but this is less than half of the 24.87% that the original STM image features.

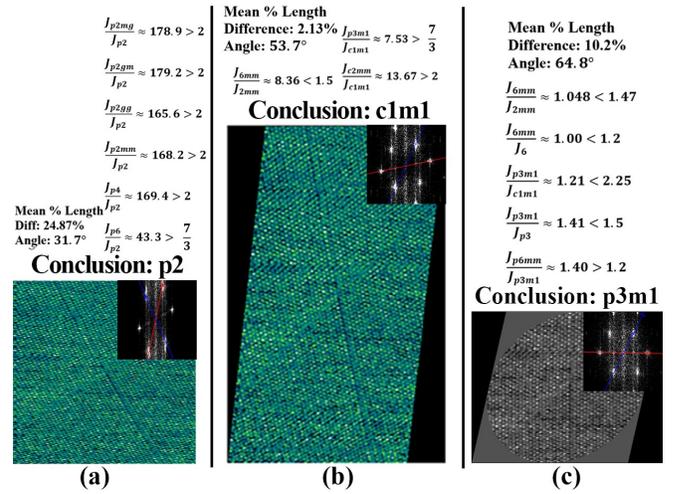

Figure 5. STM image of another HOPG sample (Test Case 2). *Pictured before (a) and after (b) correction, as well as after a second correction (c), with DFT amplitude map insets, accompanied by respective ratios of J values, a composite lattice distortion measure, and the reciprocal lattice angle. (Length scales have been dropped as atomic resolution is presented.)*

This image is shown here to demonstrate not only the efficacy and iterative nature of our correction method, but also the severity of systematic image distortions (that are often essentially invisible to the unaided human eye) in some instances. Obviously, the actual scan area was rather different from the square scan area of the image data presented to the unsuspecting operator by the STM instrument.

## C. Test case 3 and comparative analysis

The final STM image of HOPG of this paper has been taken (with permission) from a paper detailing another method of geometric distortion correction by other authors [3], Fig. 6a. Their image is rather large and visually appears quite compelling in its symmetry. Indeed, our analysis shows that their image exhibits the *p3m1* symmetry group already before its distortion correction, but with a confidence level of only 2.98%, and an inconsistent projected Laue class of *2mm*. At least the latter of these two measures indicates the presence of systematic imaging errors.

The correction performed by Yothers and co-workers [3] utilized a computationally much more demanding direct-space method that deals with distortions at the nearest neighbor level. We applied our objective symmetry classification method to their original STM image, to their corrected image, as both presented in [3], and to the results of our own correction method applied to a large central disk selection of their original image, see Fig. 6.

After the correction applied by the authors of [3], the classification remained at $p3m1$ and projected Laue class $2mm$, but with an improved confidence level of 36.65% for the plane symmetry classification. By contrast, after our correction method was applied to the original image in [3], this 2D crystal pattern yielded a $p3m1$ classification with consistent projected Laue class $6mm$ and confidence levels of 37.75% for the plane symmetry as well as 19.0% for the projected Laue class classifications, Fig. 6c.

The original crystal pattern in [3] boasted a Length Difference of 4.45% and an Angle Deviation of 2.68°. After the correction performed by Yothers and co-workers [3], the Length Difference improved to 1.55% and the Angle Deviation dropped to 0.97°, Fig. 6b. Our geometric distortion correction method, on the other hand, yielded a Length Difference of 2.24% and Angle Deviation of 1.42°, Fig. 6c.

In total we see a predominant improvement across all of our test cases. Seven of the fourteen test cases started with an objective symmetry classification of $p3m1$. Of the seven cases which exhibited other plane symmetry classifications before their correction, six were able to achieve the $p3m1$ classification after correction and one remained at plane symmetry group $c1m1$. Of the six successful symmetry corrections, one switched from $p2$ to $p3m1$ and another one switched from $p31m$ to $p3m1$. (Note that $p2$ is not a subgroup of $c1m1$ and neither is $p31m$ a subgroup of $p3m1$ [10].) Only two of the fourteen test cases started with projected Laue class $6mm$. Of the remaining twelve, six registered as such after correction. The non-negligible image distortions at the reciprocal lattice level (due to the SPM imaging processes) were reduced with a mean of 48% relative to the original images. (This number includes the "once corrected" image in Fig. 5b but not another STM image in [7] where even a second correction did not yield an improvement in interpretability, presumably as result of an excessive amount of jitter in the experimental image.)

ACKNOWLEDGMENTS

We would like to acknowledge Dr. Rama Vasudevan and Dr. Art Baddorf from the Center for Nanophase Materials Science of the Oak Ridge National Laboratory for their assistance in the gathering of STM images and Fig. 5a.

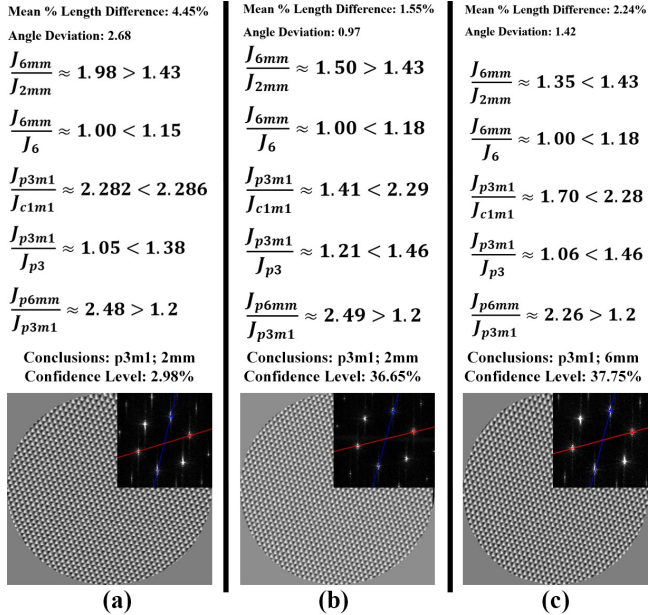

Figure 6. STM image of another HOPG sample (Test Case 3). *Pictured before (a) and after correction as performed in [3] (b) as well as after our own correction (c) with DFT amplitude map insets, accompanied by respective ratios of J values, and composite lattice distortion metrics. (Length scales have been dropped as atomic resolution is presented.)*

## VI. SUMMARY OF RESULTS ON ELEVEN MORE TEST CASES

In addition to the three test cases briefly discussed above, a synthetic image of HOPG (of our own creation) and ten other experimental atomic-resolution images from crystals that possess plane symmetry group $p3m1$ in [0001] projection have been analyzed by us. The results of each of the fourteen test cases both before and after the application of our correction method are detailed in [7].